\documentclass[preprint,aps]{revtex4-1}

\usepackage{amsmath,amssymb,amsfonts,dcolumn,color,graphicx,graphics,latexsym,placeins,epsfig}
\usepackage{epsfig,color}
\usepackage{bm}
\usepackage{latexsym}
\usepackage{natbib}
\usepackage{url}
\usepackage{dcolumn}
\usepackage{color}
\usepackage{amsfonts,amssymb,amsmath}
\usepackage{graphicx,epsfig}
\usepackage{psfrag}
\usepackage{subfigure}
\usepackage{hyperref}
\hypersetup{colorlinks=true}

\newcommand{\be}{\begin{equation}}
\newcommand{\ee}{\end{equation}}
\newcommand{\ba}{\begin{eqnarray}}
\newcommand{\ea}{\end{eqnarray}}

\begin{document}

\title{Overcharging black holes and cosmic censorship in Eddington-inspired Born-Infeld gravity }

\author{Soumya Jana$^{1,}$\footnote{sjana@prl.res.in}}
\author{Rajibul Shaikh$^{2,}$\footnote{rajibul.shaikh@tifr.res.in}}
\author{Sudipta Sarkar$^{3,}$\footnote{sudiptas@iitgn.ac.in}}

\affiliation{${}^{1}${\it Physical Research Laboratory, Ahmedabad 380009, India}}
\affiliation{${}^{2}$ {\it  Tata Institute of Fundamental Research,
Homi Bhabha Road, Colaba, Mumbai 400005, India}}
\affiliation{${}^{3}$ {\it  Indian Institute  of  Technology,  Gandhinagar 382355, India }}


\begin{abstract}

The Eddington-inspired Born-Infeld (EiBI) gravity is a modification of the theory of general relativity inspired by the nonlinear Born-Infeld electrodynamics. The theory is described by a series of higher curvature terms added to the Einstein-Hilbert action with the parameter $\kappa$. The EiBI gravity has several interesting exact neutral and charged black hole solutions. We study the problem of overcharging extremal black hole solutions of EiBI gravity using a charged test particle to create naked singularity. We show that unlike general relativity, the overcharging could be possible for a charged extremal black hole in EiBI gravity as long as the matter sector is described by usual Maxwell's electrodynamics. Once the matter sector is also modified in accordance to the Born-Infeld prescription with the parameter $b$, the overcharging is not possible as long as the parameters obey the condition $4 \kappa b^2 \leq 1$.

\end{abstract}


\maketitle

\section{\bf Introduction} 


General relativity (GR) is extremely successful as a classical theory 
of gravity and over the years, it has been under scrutiny in vacuum or in the weak-field regime through several precision tests and no significant deviation from GR has been found \cite{Will2014}. 
Still there exist many unsolved puzzles in GR such as the problem of singularities, understanding the dark matter and dark energy, etc.  
In order to address some of these problems, many researchers actively pursue 
modified gravity theories in the classical domain which deviate from GR inside matter distributions, or in the strong-field regime. One such modification is inspired by the well-known Born-Infeld electrodynamics \cite{born} where, even at the classical level, it is possible to avoid the infinity in the electric field at the location of a point charge. Deser and Gibbons \cite{desgib} first suggested a gravity theory in the metric formalism consisting a similar structure $\sqrt{-|g_{\mu\nu}+\kappa R_{\mu\nu}|}$ as in the action of Born-Infeld electrodynamics. In fact, the form of the gravitational action is not a new concept but existed earlier in Eddington's re-formulation of GR in de Sitter spacetime \cite{edd}. This is essentially an affine formalism where the affine connection is the basic variable instead of the metric, but the coupling of matter to this new formulation of gravity remained a problem.

Later, the Palatini (metric-affine) formulation in Born-Infeld gravity was introduced by Vollick \cite{vollick}. He worked on various related aspects and also introduced a nontrivial and somewhat artificial way of coupling matter in such a theory \cite{vollick2,vollick3}. More recently, Banados and Ferreira  \cite{banados} have come up with a formulation, popularly known as the Eddington-inspired Born-Infeld (EiBI) gravity, where the matter coupling is different and simpler compared to Vollick's original proposal. For a recent review on Born-Infeld gravity, see \cite{jimenez2017} and for its cosmological, astrophysical, and other applications see~\cite{cardoso,delsate,pani,scargill,cho,chokim,escamilla,linear.perturbation,wei,jana,
jana2,jana4,jana2017,jana18a,rajibul,rajibul18a,sotani,eibiwormhole,BTZ_typesoln,casanellas,avelino,sham,sham2,structure.exotic.star,sotani.neutron.star, sotani.stellar.oscillations,sotani.magnetic.star,wei,sotani,odintsov,fernandes,
latorre2017,Jimenez17,olmo1705,olmo1709,escamilla18,ramadhan17,li_2017,ramadhan18,shaikhJul18} and the references therein.

Some work also have been done on black hole physics, or, broadly on 
the spherically symmetric, static solutions in this theory. It may be noted 
that the vacuum, spherically symmetric static solution in this theory is trivially same as 
the Schwarzschild de Sitter black hole. But, the electrovacuum solutions are 
expected to deviate from the usual Reissner-Nordstr\"om solution in GR. 
This has been shown in \cite{banados,wei,sotani} where the authors 
consider EiBI gravity coupled to a Maxwell electric field of a 
localized charge. They obtain the resulting spacetime geometries, and study its properties. The basic features of such spacetimes includes a singularity 
at the location of the charge which may or may not be covered by an 
event horizon. The strength of the electric field remains 
nonsingular as in Born-Infeld electrodynamics. 
However, this may not be the only solution because, in EiBI gravity, 
the matter coupling is nonlinear. In a different framework \cite{eibiwormhole}, it was shown that the central singularity could be 
replaced by a wormhole supported by the electric field. In \cite{rajibul}, the author obtained a class of Lorentzian regular wormhole spacetimes supported by the quintessential matter which does not violate the weak or null energy condition in EiBI gravity. The generalisation of this result in the context of arbitrary nonlinear
electrodynamics and anisotropic fluids was obtained in \cite{shaikhJul18}. Some new classes of spherically symmetric static spacetimes were obtained where EiBI gravity is coupled with Born-Infeld electrodynamics \cite{jana2}. They include black holes and naked 
singularities. Earlier, a lot of work had indeed been done by considering 
nonlinear electrodynamics coupled to GR \cite{geon,Einstein.BI.1,Einstein.BI.2,Einstein.BI.3,dibakar1,dibakar2,dibakar3,jonas}. Some of them were motivated by string theory since 
Born-Infeld structures naturally arise in the low energy limit of 
open string theory \cite{openstring1,openstring2}.

An essential question in general relativity is to understand the global properties of the field equation, in particular, the issue of cosmic censorship. There are various versions of cosmic censorship conjecture. One of the version prohibits evolution of a generic, sufficiently regular initial data into a solution with a naked singularity. The full analysis of this problem is complicated given the complicated nature of the Einstein's field equations. A more straightforward exercise could be to look for specific counterexamples, where one starts with a black hole solution with a horizon and try to create a naked singularity using a physical process. For example, Wald \cite{wald1974} considered the problem of overcharging an extremal Reissner-Nordstr\"om (R-N) black hole solution using a charged test particle. Interestingly, the dynamics of the particle does not allow such overcharging to happen. In \cite{hubeny}, the problem was studied for a near extremal R-N black hole. It was shown that overcharging is possible if the back-reaction effects are ignored. Similar consideration was obtained from the study of the rotating black hole in \cite{Ted} and also for massless charged particles \cite{Sudipta}. The back reaction problem was analyzed in detail in \cite{Zimmerman} and it was shown that the overcharging would not occur once the back-reaction effects are considered. In the context of general relativity, a general proof of the impossibility of overcharging an extremal or near-extremal black hole solution was provided in \cite{wald_new} generalizing a result in \cite{Natario}. Related works had also been done for the black holes in higher dimensional gravity \cite{vega_2017,an_2018,ge_2018,dadhich_2018}. 

In this work, we study the same overcharging problem in the context of EiBI gravity. We analyze the dynamics of charged test particles in the background of extremal black hole solutions in the EiBI gravity and show that the overcharging could be possible when the matter sector is described by usual Maxwell's electrodynamics. Interestingly, once we consider the modification of the matter sector by the Born-Infeld prescription, we find that there is no possibility of overcharging (provided a condition on the Born-Infeld parameters is satisfied). Our result indicates that the Born-Infeld modification of gravity along with matter sector is as consistent as general relativity.

\section{Overcharging a black hole by throwing a massive charged particle}
We consider the motion of a test particle of charge $q$, mass $m$, and four-velocity $u^{\mu}$, in a fixed background spacetime (spherically symmetric and static) given by
\begin{equation}
ds^2=g_{tt}(r)dt^2+g_{rr}(r)dr^2+r^2(d\theta^2+\sin^2\theta d\phi^2),
\label{eq:sphe_space}
\end{equation} 
where $g_{tt}(r)$ and $g_{rr}(r)$ are characterized by the black hole parameters: charge $Q$, mass M, and the Born-Infeld parameters $\kappa$ and $b^2$ (to be introduced later).
The motion of the test particle can be obtained from the following Lagrangian,
\begin{equation}
L=\frac{1}{2}mu^{\mu}u_{\mu}+qu^{\mu}A_{\mu},
\label{eq:Lagrangian}
\end{equation}
where $A_{\mu}(x_{\nu})$ is the electromagnetic vector potential of the black hole. 
For radial motion of the charged particle, $u^{\mu}=\lbrace \dot{t}, \dot{r}, 0, 0 \rbrace$, $\dot{t}=\frac{dt}{d\lambda}$, $\dot{r}=\frac{dr}{d\lambda}$, $\lambda$ being the affine parameter along the world line. Then, from Eqs.~(\ref{eq:sphe_space}) and (\ref{eq:Lagrangian}) we get 
\begin{equation}
\frac{\partial L}{\partial \dot{t}}= mg_{tt} \dot{t} + qA_t= -E,
\label{eq:E}
\end{equation}
where $E$ is a constant of motion along the particle's worldline. Then, for the timelike trajectories, i.e. $u^{\mu}u_{\mu}=-1$,
\begin{equation}
\dot{r}^2= -\frac{(E+qA_t)^2}{m^2 g_{tt}g_{rr}}-\frac{1}{g_{rr}}
\label{eq:rdot}
\end{equation}
For the Reissner-Nordstr\"om (R-N) black hole solution $A_t=-Q/r$. However, $A_t$ is modified in the presence of the Born-Infeld structures in gravity and matter sectors.
Since $\dot{r}=0$ corresponds to a turning point, for ``in fall" of the particle
\begin{equation}
\dot{r}^2 > 0 ,~\quad \mbox{for all}~\, r\geq r_+,
\label{eq:fall_in}
\end{equation}
where $r_+$ is the event horizon corresponding to the initial configuration of the black hole. When the particle falls past the radial coordinate $r_+$, the final configuration of the black hole consisting of total charge $(Q+q)$ and mass $(M+E)$  must exceeds extremality in order to destroy the black hole. In case of the R-N black hole solution, this implies that, $Q+q>M+E$. 

In \cite{wald1974}, it is established that these two conditions are mutually exclusive and can not be satisfied together. As a result, it is impossible to overcharge an extremal charged black hole in GR to create a naked singularity.

In Born-Infeld theories, the charged black hole solution is modified and the condition of overcharging becomes,

\begin{equation} \label{eq:exceeding_extreem}
M+E< \overline{Q+q}
\end{equation}
where $\bar{Q}\equiv \bar{Q}(Q,\kappa,b^2)$ is an ``effective charge" and is a function of the actual black hole charge $Q$ and the BI parameters $\kappa$ and $b^2$. Thus in equation~(\ref{eq:exceeding_extreem}), $\overline{Q+q}=\bar{Q}(Q+q,\kappa,b^2)$. For the initial extremal black hole, we have $\bar{Q}(Q,\kappa,b^2)=M$. In the R-N limit, i.e. $\kappa\rightarrow 0$ and $b^2\rightarrow \infty$, $\bar{Q}(Q)=Q$ and $\bar{Q}(Q+q)=Q+q$. We assume the ``back reaction" effects are negligible. Therefore to overcharge a black hole, the two conditions given by Eqs.~(\ref{eq:fall_in}) and (\ref{eq:exceeding_extreem}) must be satisfied.\\

\section{The Eddington-inspired Born-Infeld (EiBI) gravity}
First we briefly recall the details of EiBI gravity. The action for the theory developed in Ref. \cite{banados} is given as
\ba
S_{EiBI}(g_{\mu\nu},\Gamma, \Psi) &=&\frac{c^3}{8\pi G\kappa}\int d^4 x \left[\sqrt{-\vert g_{\mu\nu} +\kappa R_{\mu\nu}(\Gamma)\vert} -\lambda \sqrt{-\vert g_{\mu\nu}\vert} \right]+ S_M (g_{\mu\nu}, \Psi)
\label{eq:eibi_action}
\ea
where $S_M(g_{\mu\nu},\Psi)$ is the matter part of the full gravitational action, $\Psi$ generically denotes any matter field, and $\kappa$ is the constant parameter of the theory having dimension of [Length]${}^2$. $R_{\mu\nu}$ is assumed to be the symmetric part of the Ricci tensor constructed from the connection $\Gamma$.  In Einstein's limit, i.e. for $\kappa R_{\mu\nu}<< g_{\mu\nu} $, the action reduces to the Einstein-Hilbert action, provided the dimensionless parameter $\lambda$ corresponds to the cosmological constant $\Lambda$ as $\lambda \equiv \kappa \Lambda +1$. The theory is based on the Palatini formulation, and therefore, the metric ($g_{\mu\nu}$) and the connection ($\Gamma^{\rho}_{\mu\nu}$) are treated as independent variables in the action. By varying~(\ref{eq:eibi_action}) with respect to $\Gamma^{\alpha}_{\mu\nu}$, one obtains
\begin{eqnarray}
{\nabla^{\Gamma}}_{\alpha}(\sqrt{-q}q^{\mu\nu})=0,\label{eq:eibif1}\\
\mbox{where}\quad~ q_{\mu\nu}\equiv  g_{\mu\nu}+\kappa R_{\mu\nu}(\Gamma),
\label{eq:eibif1b}
\end{eqnarray}
 $q=\det(q_{\mu\nu})=\vert g_{\mu\nu}+\kappa R_{\mu\nu}\vert$, and $\nabla^{\Gamma}$ denotes the covariant derivative w.r.t. the connection $\Gamma$.  Equation~(\ref{eq:eibif1}) shows that  $\Gamma$ is compatible w.r.t. the metric $q_{\mu\nu}$, and hence, we can compute $\Gamma$ using the equation
\be   
\Gamma^{\mu}_{\;\alpha\beta} =\frac{1}{2}q^{\mu\sigma}\left( q_{\sigma\alpha,\beta}+q_{\sigma\beta,\alpha}-q_{\alpha\beta,\sigma}\right)\; .
\label{eq:eibif2}
\ee
$q^{\mu\nu}$ is the inverse  $q$-metric such that $q^{\mu\alpha}q_{\mu\beta}=\delta^{\alpha}_{\beta}$.
The matter field is minimally coupled only to $g_{\mu\nu}$, i.e. the matter part of the full gravitational action ($S_M (g_{\mu\nu}, \Psi)=\frac{1}{c}\int \sqrt{-g}L_M(g_{\mu\nu},\Psi)d^4x $) only depends on the metric $g_{\mu\nu}$ and on the matter field $\Psi$, but not on the connection $\Gamma$. Therefore, the metric $g_{\mu\nu}$ is the physical metric, whereas the metric $q_{\mu\nu}$ is called as the auxiliary metric. Thus, the connection ($\Gamma$) is different from the Levi-Civita connection ($\lbrace {}\rbrace^{\mu}_{\;\alpha\beta} =\frac{1}{2}g^{\mu\sigma}\left(g_{\sigma\alpha,\beta}+g_{\sigma\beta,\alpha}-g_{\alpha\beta,\sigma}\right)$).

Variation of the action~(\ref{eq:eibi_action}) with respect to $g_{\mu\nu}$ gives 
\begin{eqnarray}
\sqrt{-q}q^{\mu\nu}&=&\lambda\sqrt{-g}g^{\mu\nu}-\frac{8\pi G\kappa}{c^4} \sqrt{-g}\; T^{\mu\nu} ,\label{eq:eibif3}
\end{eqnarray}
where  $g=\det(g_{\mu\nu})=\vert g_{\mu\nu}\vert$ and $T^{\mu\nu}=\frac{2}{\sqrt{-g}}\frac{\partial \left(\sqrt{-g}L_{M}\right)}{\partial g_{\mu\nu}}$ are components of the stress-energy tensor in the coordinate frame. The stress-energy tensor is conserved ($\nabla_{\mu}T^{\mu\nu}=0$) w.r.t. the physical metric $g_{\mu\nu}$. The index of $T_{\mu\nu}$ and $R_{\mu\nu}(\Gamma)$ are to be raised/lowered by $g_{\mu\nu}$ and $q_{\mu\nu}$ respectively. Note that Eq.~(\ref{eq:eibif3}) is just an algebraic equation relating the physical metric to the auxiliary metric through the stress-energy tensor. Equation~ (\ref{eq:eibif1}) (along with Eq.~(\ref{eq:eibif2})) gives a set of differential equations which are to be solved simultaneously with Eq.~(\ref{eq:eibif3}) to get the full solutions. Therefore, Eqs.~(\ref{eq:eibif1b})-(\ref{eq:eibif3}) constitute the gravitational field equations in EiBI theory.

The structure of EiBI theory implies that the physical metric ($g_{\mu\nu}$) governs the dynamics of test particles. In more precise words, a freely falling test particle follows the geodesics of the physical spacetime. Therefore, invariant scalar quantities associated with the physical spacetime metric such as $R(g_{\mu\nu}), R_{\alpha\beta}^2(g_{\mu\nu})$ etc. are relevant. On the other hand, the auxiliary metric ($q_{\mu\nu}$) is introduced in the field equations for mathematical convenience. It does not couple to matter fields but plays an indirect role through its presence in the field equations.

\section{Black holes supported by the Maxwell's electric field and the overcharging problem}

For the Maxwell's electromagnetic field theory in the curved spacetime, the Lagrangian density is $\mathcal{L}=-\frac{1}{16\pi}\sqrt{-g}F_{\mu\nu}F^{\mu\nu}$, where $F_{\mu\nu}=\partial_{\mu}A_{\nu}-\partial_{\nu}A_{\mu}$ is the electromagnetic field tensor. The corresponding stress-energy tensor is given by $T_{\mu\nu}=-\frac{2}{\sqrt{-g}}\frac{\partial \mathcal{L}}{\partial g^{\mu\nu}}=\frac{1}{4\pi}\left(F_{\mu\sigma}
F_{\nu}{}^{\sigma}-\frac{1}{4}g_{\mu\nu}F_{\alpha\beta}F^{\alpha\beta}\right)$. For an electrostatic scenario, the four-potential is $A_{\mu}=\lbrace A_t(r),0,0,0\rbrace$.

\subsection{General relativity}

In GR, i.e. for the Reissner-Nordstr\"om spacetimes, $g_{tt}=-1/g_{rr}=-\left(1-\frac{2M}{r}+\frac{Q^2}{r^2}\right)$, $A_t= -Q/r$. The event horizon ($r_+$) and the Cauchy horizon ($r_{-}$) are given by $r_{\pm}=M\pm \sqrt{M^2-Q^2}$. The extremality corresponds to $r_{+}=r_{-}=Q=M$. One can also note that, for extremality, $g_{tt}(r_{e})=g_{tt}'(r_e)=0$, where the extremal horizon radius $r_e=r_{+}=r_{-}$. Then, from
Eqs.~(\ref{eq:rdot}) and (\ref{eq:fall_in}), we get that $E>q$ for the test particle falling past the horizon of the extremal black hole. On the other hand, to exceed the extremality condition of the final black hole, we need $E<q$. So, there is no window of choosing a suitable $E$. Thus, the overcharging is not possible for the Reissner-Nordstr\"om extremal black hole by throwing a massive charged particle. This is the result obtained in \cite{wald1974}.

\subsection{EiBI gravity}

\begin{figure}[!htbp]
  \centering
    \includegraphics[width=3.0in,angle=360]{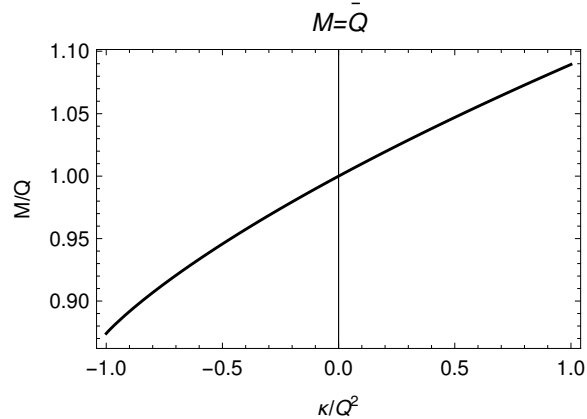}
   \caption{Plot of the mass to charge ratio ($M/Q$) of the extremal EiBI-Maxwell black holes (i.e. for $M=\bar{Q}$) as the function of $\kappa/Q^2$ using Eq.~(\ref{eq:extremal_mass1}). We use the restriction $\vert \kappa/Q^2\vert < 1$ in the plot. }
\label{fig:MbyQ}
\end{figure}

\begin{figure}[!htbp]
\centering
\subfigure[$\kappa/Q^2= 0.01$]{\includegraphics[width=3.0in,angle=360]{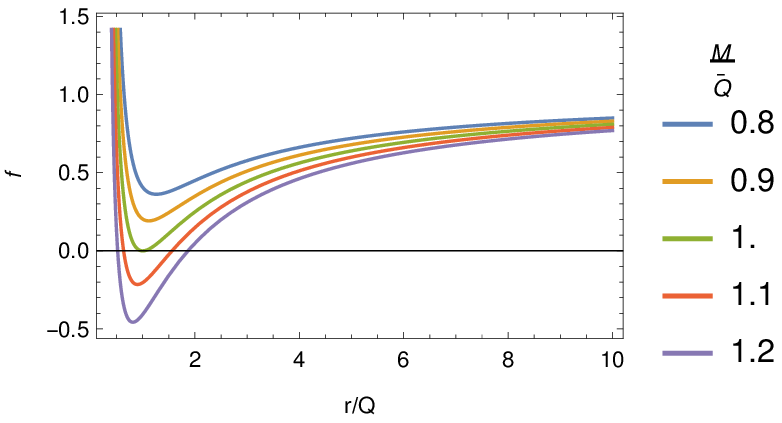}\label{subfig:f1}}
\subfigure[$\kappa/Q^2= 0.5$]{\includegraphics[width=3.0in,angle=360]{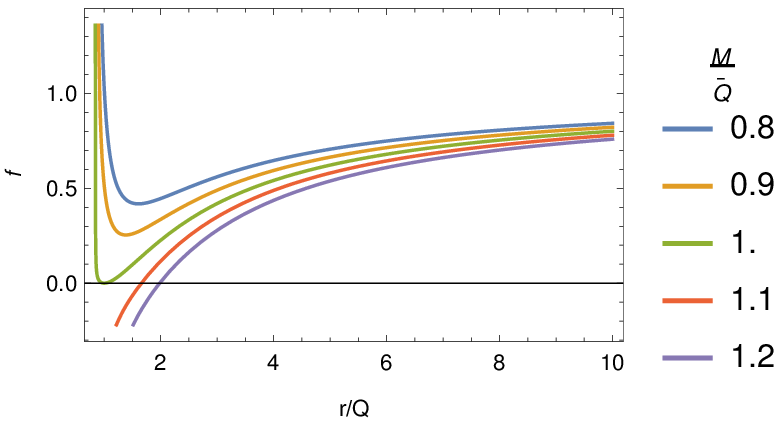}\label{subfig:f3}}
\subfigure[$\kappa/Q^2= -0.01$]{\includegraphics[width=3.0in,angle=360]{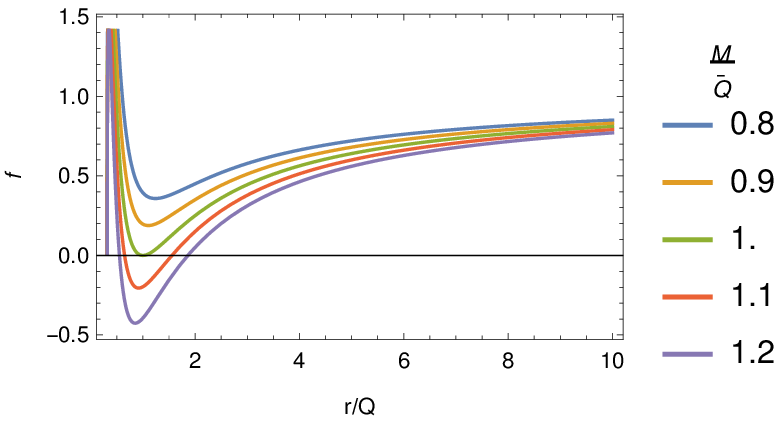}\label{subfig:f2}}
\subfigure[$\kappa/Q^2= -0.1$]{\includegraphics[width=3.0in,angle=360]{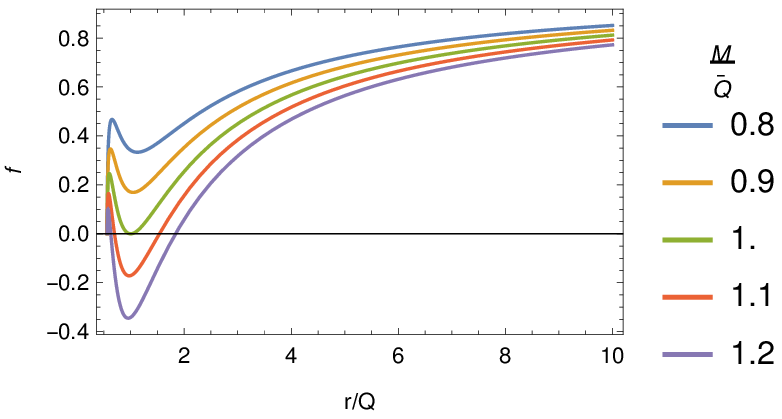}\label{subfig:f4}}
\caption{At the horizon $f(r)=0$. For black holes, there must be at least one zero of $f(r)$, and for naked singularities $f(r)>0$. Here we plot $f$ (given by Eq.~(\ref{eq:f})) as the function of $r/Q$. Given a fixed value of $\kappa/Q^2$ ($\kappa/Q^2=0.01, 0.5, -0.01, -0.1$ in the plots (a), (b), (c), and (d) respectively) we plot $f(r)$ for several values of $M/\bar{Q}$ ($M/\bar{Q}=0.8, 0.9, 1, 1.1, 1.2$) and compare. In each of the plots (a)-(d), we note that $M=\bar{Q}$ is the critical condition which distinguishes the black holes from naked singularities. Clearly, we have black holes for $M\geq \bar{Q}$ and naked singularities for $M< \bar{Q}$. Therefore, $M=\bar{Q}$ (given in Eq.~(\ref{eq:extremal_mass1})) is the extremal condition of EiBI black holes. This is similar to the case of GR where we have black holes if $M\geq Q$ and naked singularities if $M<Q$. }
\label{fig:extremal}
\end{figure}

In EiBI gravity, the resulting black hole spacetime is given by \cite{banados,wei,sotani,rajibul,jana2,shaikhJul18} $g_{tt}=-\psi^2(r)f(r)$ and $g_{rr}=1/f(r)$, where
\begin{eqnarray}
\psi &=& \left[1+\frac{\kappa Q^2}{r^4}\right]^{-1/2},\label{eq:psi}\\
f(r) &=& \left(\frac{1+\frac{\kappa Q^2}{r^4}}{1-\frac{\kappa Q^2}{r^4}}\right)\left[1-\frac{2M}{r\sqrt{1+\frac{\kappa Q^2}{r^4}}}-\frac{Q^2}{3r^2}+\frac{4Q^2}{3r^2\sqrt{1+\frac{\kappa Q^2}{r^4}}}\,{}_2F_1\left(\frac{1}{2},\frac{1}{4};\frac{5}{4};-\frac{\kappa Q^2}{r^4}\right)\right] .
\label{eq:f}
\end{eqnarray}     
By solving the equation of motion for the electric scalar potential $A_t$, or alternatively from the conservation of the stress-energy tensor (i.e. $\nabla_{\mu}T^{\mu\nu}=0$) we get
\begin{equation}
A_t=\int\frac{Q dr}{r^2\sqrt{1+\frac{\kappa Q^2}{r^4}}}= -\frac{Q}{r}\,{}_2F_1\left(\frac{1}{2},\frac{1}{4};\frac{5}{4};-\frac{\kappa Q^2}{r^4}\right).
\label{eq:A_t1}
\end{equation} 
Note that the spacetime is singular at $r_0=(\kappa Q^2)^{1/4}$ ($(\vert \kappa \vert Q^2)^{1/4}$ for $\kappa<0$) unlike in the case of R-N black hole where we get a point singularity at $r_0=0$ and the charge $Q$ is now distributed over a 2-sphere of area radius $r_0$, instead of being a `point charge'. The horizon radius ($r_e$) of the extremal black hole is obtained from $f(r_e)=f'(r_e)=0$ using Eq.~(\ref{eq:f}). This leads to
\begin{eqnarray}
r_e &=& Q, \label{eq:extremal_horizon1}\\
\mbox{and}~\quad M&=&\frac{Q}{3}\left[\sqrt{1+\frac{\kappa}{Q^2}}+2\,{}_2F_1\left(\frac{1}{2},\frac{1}{4};\frac{5}{4};-\frac{\kappa}{Q^2}\right)\right]=\bar{Q}(Q,\kappa).
\label{eq:extremal_mass1}
\end{eqnarray}  
Thus, for extremal black holes, we have $M=\bar{Q}$ where $\bar{Q}$ is an ``effective charge" and is function of the actual charge $Q$ and BI parameter $\kappa$. Note that for $\kappa=0$ (i.e. the GR limit) in the last equation, $M=Q$. However, for $\kappa \neq 0$, the mass to charge ratio ($M/Q$) differs from $1$ (see Fig.~\ref{fig:MbyQ}). Also note that, for $r_0\geq Q$ i.e. $\vert \kappa \vert\geq Q^2$, horizon lies below $r_0$ and we do not have an extremal black hole at all. Thus we assume $\vert \kappa\vert< Q^2$ in our study. Here, for $\bar{Q}>M$, we have naked singularities similar to the case in GR for $Q>M$. We verify this by a graphical analysis shown in Fig.~\ref{fig:extremal} as it is difficult to verify analytically due to the complexity of functional form of $f(r)$ (Eq.~\ref{eq:f}). 

Using Eqs.~(\ref{eq:psi}), (\ref{eq:f}), (\ref{eq:A_t1}) in Eq.~(\ref{eq:rdot}), we get

\begin{equation}
\dot{r}^2 = \frac{1}{m^2\psi^2(r)}\left[E-\frac{qQ}{r}\,{}_2F_{1}\left(\frac{1}{2},\frac{1}{4};\frac{5}{4};-\frac{\kappa Q^2}{r^4}\right)\right]^2-f(r).
\label{eq:rdot1}
\end{equation}
Then, for crossing the horizon, $\dot{r}^2>0$ for all $r\geq r_e$. To satisfy this condition at the horizon radius, $r=r_e=Q$,
\begin{equation}
E> q\cdot\,{}_2F_1\left(\frac{1}{2},\frac{1}{4};\frac{5}{4};-\frac{\kappa}{Q^2}\right).
\label{eq:infall1}
\end{equation}
We use Eq.~(\ref{eq:exceeding_extreem}) to get the condition for exceeding the extremality of the final black hole
\begin{equation}
E< \bar{Q}(Q+q,\kappa) -\bar{Q}(Q,\kappa),
\label{eq:exceed_extreem1}
\end{equation}
where we used $M=\bar{Q}(Q,\kappa)$ for the initial extremal black hole configuration and $\bar{Q}(Q,\kappa)$ is given by Eq.~(\ref{eq:extremal_mass1}).

Both Eqs. (\ref{eq:infall1}) and (\ref{eq:exceed_extreem1}) will be simultaneously satisfied, i.e., there will be a window for a choice of $E$ for overcharging the black hole only when the quantity
\begin{equation}
\Delta=\bar{Q}(Q+q,\kappa) -\bar{Q}(Q,\kappa) -q\cdot{}_2F_1\left(\frac{1}{2},\frac{1}{4};\frac{5}{4};-\frac{\kappa}{Q^2}\right)
\end{equation}
is positive ($\Delta>0$). However, in general, showing $\Delta>0$ analytically is difficult. For small $\kappa$ or for large black hole such that $Q^2\gg |\kappa|$, we obtain
\begin{eqnarray}
{}_2F_1\left(\frac{1}{2},\frac{1}{4};\frac{5}{4};-\frac{\kappa}{Q^2}\right)&\simeq & 1-\frac{\kappa}{10 Q^2}\\
\mbox{and}~\quad~ \bar{Q}(Q,\kappa)&\simeq & Q+\frac{\kappa}{10Q}.
\end{eqnarray}
Also, noting that the test charge $q$ must be small compared to the black hole charge $Q$, i.e. $Q\gg q$, we obtain
\begin{equation}
\Delta\simeq \frac{\kappa q^2}{10Q^2(Q+q)}\simeq \frac{\kappa q^2}{10Q^3}.
\end{equation}
Hence, overcharging of the extremal black hole is always possible for $\kappa>0$. However, for $\kappa\leq 0$, overcharging is not possible. Also, note that we recover the general relativistic results in the limit $\kappa\to 0$.

\begin{figure}[!htbp]
\centering
\subfigure[$\kappa>0$]{\includegraphics[width=3.0in,angle=360]{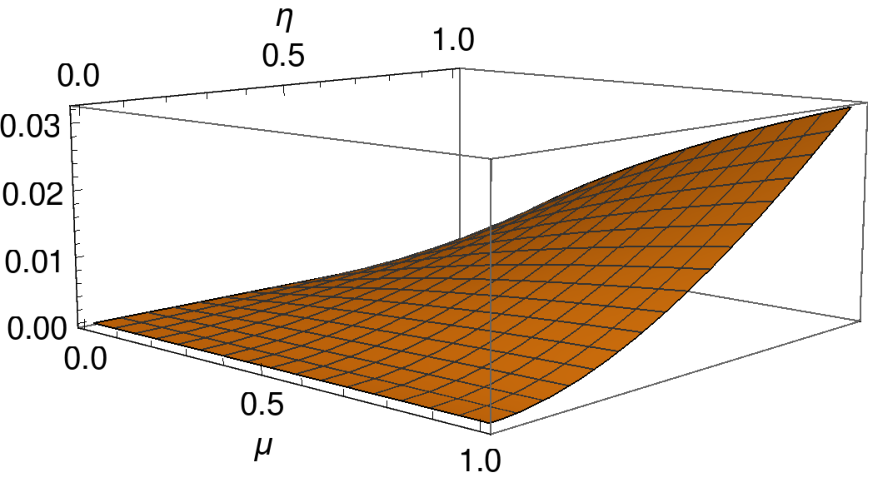}\label{subfig:eibi_maxwell1}}
\subfigure[$\kappa<0$]{\includegraphics[width=3.0in,angle=360]{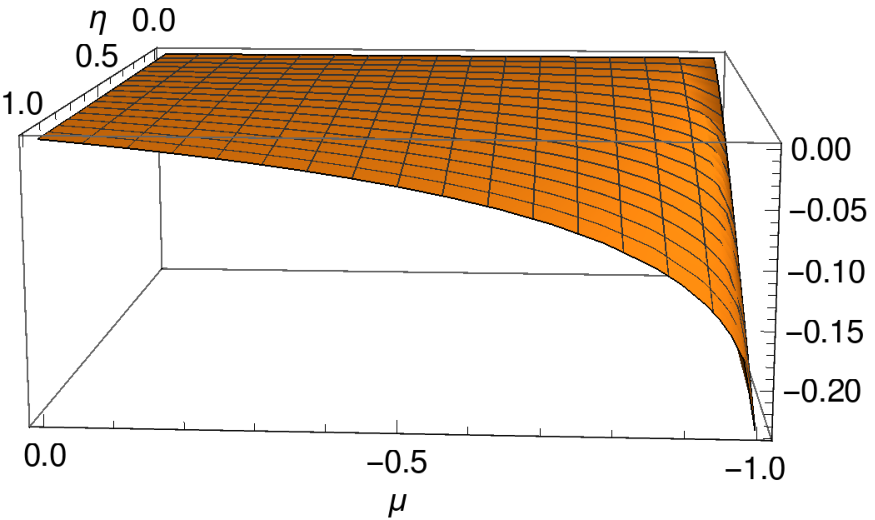}\label{subfig:eibi_maxwell2}}
\caption{Surface plot of $\xi(\mu, \eta)$ for $\vert \kappa \vert \leq Q^2$ and $q\leq Q$. $\mu$ and $\eta$ are dimensionless variables defined as $\mu=\frac{\kappa}{Q^2}$ and $\eta=\frac{q}{Q}$. $\xi>0$ for all $\mu>0$ (i.e. $\kappa>0$) and $\eta$ (i.e. $q$) where as $\xi<0$ for all $\mu<0$ and $\eta$. Thus, overcharging an extremal black hole is possible for all $\kappa>0$ and for small $q$ (smaller than the black hole charge Q).}
\label{fig:eibi_maxwell}
\end{figure}

To show whether the overcharging is possible for arbitrary $\kappa$ and $Q$, we define a dimensionless variable $\xi(\mu,\eta)$, (where $\mu=\frac{\kappa}{Q^2}$ and $\eta=\frac{q}{Q}$), as
\begin{eqnarray}
\xi &=& \frac{\Delta}{Q}= \frac{(1+\eta)}{3}\left[\sqrt{1+\frac{\mu}{(1+\eta)^2}}+2\,{}_2F_1\left(\frac{1}{2},\frac{1}{4};\frac{5}{4};-\frac{\mu}{(1+\eta)^2}\right)\right]\nonumber\\
&&-\frac{1}{3}\left[\sqrt{1+\mu}+2\,{}_2F_1\left(\frac{1}{2},\frac{1}{4};\frac{5}{4};-\mu\right)\right]-\eta\,{}_2F_1\left(\frac{1}{2},\frac{1}{4};\frac{5}{4};-\mu\right).
\label{eq:xi}
\end{eqnarray}
If $\xi>0$ for some specific values of $\mu$ and $\eta$, then there will be a window for a choice of $E$ for overcharging the black hole. In Fig.~\ref{fig:eibi_maxwell}, we plot (3D surface plot) $\xi$ as the function of $\mu$ and $\eta$ where we use $\vert \kappa \vert \leq Q^2$ (i.e. $\vert\mu\vert \leq 1$) and  $q\leq Q $ (i.e. $\eta\leq 1$). From the plot, we note that $\xi>0$ for all $\mu>0$ and $\eta$.  Thus, for $\kappa>0$, we can choose the energy $E$ of the test particle with any small charge $q$ (smaller than the black hole charge $Q$), so that the inequalities Eqs.~(\ref{eq:infall1}) and (\ref{eq:exceed_extreem1}) will be satisfied. 
   
For in-falling of the test particle, $\dot{r}^2>0$ for all $r>r_e$. This implies that (using Eq.~(\ref{eq:rdot1}))
\begin{equation}
m < \frac{E-\frac{q \, Q \,{}_2F_1\left(\frac{1}{2},\frac{1}{4};\frac{5}{4};-\frac{\kappa Q^2}{r^4}\right)}{r}}{\psi(r)\sqrt{f(r)}}, \, r\geq r_e.
\label{eq:m}
\end{equation}
R.H.S. of the inequality (\ref{eq:m}) is monotonically decreasing and reaches the value $E$ asymptotically at large $r$. Therefore, inequality (\ref{eq:m}) is satisfied for $m<E$ which is true for any ordinary matter.
 
Thus, for $\kappa>0$, the overcharging an extremal black hole is always possible by throwing a test charged particle of small charge $q$ and energy $E$ satisfying the conditions Eqs.~(\ref{eq:infall1}) and (\ref{eq:exceed_extreem1}). This is a significant departure from the result obtained in the case of GR, where even without the back-reaction, it is not possible to create a naked singularity by overcharging an extremal charged black hole. The test charge required for such a process can never enter the black hole. But, in EiBI theory, since the dynamics are different, it is possible to find a situation where overcharging an extremal black hole is possible. If we can create a naked singularity from an extremal solution using a physical process, it is a counterexample to the cosmic censorship in the context of EiBI gravity. It seems unlike GR, it is easier to invalidate cosmic censorship for Born-Infeld modification of the gravity.

Next, we would like to know if this can be avoided, provided we modify the matter section also using the Born-Infeld prescription. In the next section, we will study the overcharging problem for Black holes supported by the Born-Infeld electric field.

\section{Black holes supported by the Born-Infeld electric field and the overcharging problem}
 A nonlinear theory of electrodynamics was proposed by Born and Infeld in 1934 \cite{born}. In Maxwell's theory of electrodynamics, singularities appear in the electric and magnetic fields. As an example, the electric field as well as the self-energy for a point charge diverge at its location. Born and Infeld, in their theory, introduced a new parameter $b$ which sets a maximum limit on the value of the electromagnetic (EM) field, similar to the maximum speed limit in the special theory of relativity. In curved spacetime, the Lagrangian density for the BI EM field theory is given by \cite{born}
\begin{equation}
\mathcal{L}_{BI}=\frac{b^2\sqrt{-g}}{4\pi}\left[1-\sqrt{1+\frac{F}{b^2}-\frac{\mathcal{G}^2}{b^4}}\right],
\label{eq:BIED_lagrangian}
\end{equation} 
where $F=\frac{1}{2}F_{\mu\nu}F^{\mu\nu}$ and 
$\mathcal{G}=\frac{1}{4}F_{\mu\nu}\mathcal{G}^{\mu\nu}$ are two scalar 
quantities constructed from the components of the EM field 
tensor ($F_{\mu\nu}$) and the dual field tensor ($\mathcal{G}^{\mu\nu}=\frac{1}{2\sqrt{-g}}\epsilon^{\mu\nu\alpha\beta}F_{\alpha\beta}$). For an electrostatic scenario  in flat Minkowski spacetime, $\mathcal{G}=\vec{E}\cdot \vec{B}=0$, and therefore, the above Lagrangian reduces to $\mathcal{L}_{BI}=\frac{b^2}{4\pi}\left[1-\sqrt{1-\vert \vec{E}\vert^2/b^2}\right]$, where $\vec{E}$ and $\vec{B}$ are the electric and magnetic field vectors. Here, it is clear that $b$ sets an upper limit on the electric field, and consequently the self-energy is also finite for a point charge. Consequently, the Coulomb's law in BI theory gets altered as $E(r)= \frac{Q}{\sqrt{r^4+Q^2/b^2}}$. Maxwell's theory is recovered in the limit $b\rightarrow \infty$. Note that singularities in the classical EM fields are well resolved in the Quantum Electrodynamics (QED) theory which is extremely successful. However, at the time of proposal of BI electrodynamics, there was no full quantum theory of electrodynamics. BI theory was almost totally forgotten for a long time after QED came. However, recently, there is a new interest in BI theory due to investigations in string theory \cite{openstring1,openstring2}. 


The energy-momentum tensor associated with BI EM fields has the general expression
\begin{equation}
T_{\mu\nu}=-\frac{2}{\sqrt{-g}}\frac{\partial \mathcal{L}}{\partial g^{\mu\nu}}=-\frac{b^2}{4\pi}\left[g_{\mu\nu}\left(\sqrt{1+\frac{F}{b^2}-\frac{\mathcal{G}^2}{b^4}}-1\right)-\frac{b^2F_{\mu\sigma}F^{\, \sigma}_{\nu}-\mathcal{G}^2g_{\mu\nu}}{b^4\sqrt{1+\frac{F}{b^2}-\frac{\mathcal{G}^2}{b^4}}}\right],
\label{eq:stress-tensor-BI}
\end{equation}  
which is obtained from the Lagrangian~(\ref{eq:BIED_lagrangian}). Note that BI electrodynamics is a gauge invariant theory, and therefore, the Lorentz force equations are still valid for the motion of a test charged particle in the BI EM fields. However, the test particle feels a different strength of the EM force. 

\subsection{General relativity}
In GR, the black hole solution with 
Born-Infeld electric field due to a point charge is known as geonic black hole solution \cite{geon}. In this scenario, a distant observer associates a total mass which comprises $M$ (the black hole mass) and a pure electromagnetic mass stored as the 
self energy in the electromagnetic field. If $M$ is zero, the spacetime becomes regular everywhere. The spacetime for such a geonic black hole is given by $g_{tt}=-1/g_{rr}=-g_e(r)$ where
\begin{equation}
g_e(r)= 1-\frac{2M}{r}-\frac{2Q^2}{3\left(\sqrt{r^4+Q^2/b^2}+r^2\right)}+\frac{4Q^2}{3r^2}\, {}_{2}F_1\left(\frac{1}{4}, \frac{1}{2}; \frac{5}{4}; -\frac{ Q^2}{b^2r^4}\right). \label{eq:ge}
\end{equation}
By solving the equation of motion for the potential $A_t$, or alternatively from the conservation of the stress-energy tensor given in Eq.~(\ref{eq:stress-tensor-BI}) (i.e. $\nabla_{\mu}T^{\mu\nu}=0$), we get
\begin{equation}
A_t=\int\frac{Q dr}{r^2\sqrt{1+\frac{ Q^2}{b^2 r^4}}}= -\frac{Q}{r}\,{}_2F_1\left(\frac{1}{2},\frac{1}{4};\frac{5}{4};-\frac{ Q^2}{b^2 r^4}\right).
\label{eq:A_t2}
\end{equation} 

Note that the potentials $A_t$ in Eqs.~(\ref{eq:A_t1}) and (\ref{eq:A_t2}) look identical provided we define $b^2\equiv 1/\kappa$ ($\kappa>0$). This may be due to the fact that, as shown in Ref.~\cite{eibi_maxwell_map}, the EiBI
gravity coupled with Maxwell's electrodynamics can be mapped to GR coupled with BI electrodynamics. However, the Einstein equation in the mapped general relativity is that of the auxiliary metric ($q_{\mu\nu}$), but not of the physical metric ($g_{\mu\nu}$). Therefore, the physical metric in the two cases (i.e., EiBI gravity coupled to Maxwell electrodynamics and GR coupled to BI electrodynamics) will be different from one another. Also note that, although the two electromagnetic potential look identical, the corresponding physical quantities such as the energy density, pressures which appear in the field equations are different in the two cases.

There is a point singularity at $r_0=0$. For the extremal configuration, the horizon radius $r_e$ and the relation between the black hole charge $Q$ and mass $M$ become (using $g_e(r_e)=g_e'(r_e)=0$)
\begin{eqnarray}
r_e&=& Q\sqrt{1-\frac{1}{4b^2Q^2}},
\label{eq:extremal_horizon2}\\
\mbox{and}~\quad M&=&\frac{Q}{3}\left[\sqrt{1-\frac{1}{4b^2 Q^2}}+\frac{2}{\sqrt{1-\frac{1}{4b^2 Q^2}}}\,{}_2F_1\left(\frac{1}{2},\frac{1}{4};\frac{5}{4};-\frac{ Q^2}{b^2 r^4_e}\right)\right]\nonumber\\
&=&\bar{Q}(Q,b^2).
\label{eq:extremal_mass2}
\end{eqnarray}
Note that $4b^2Q^2> 1$ for the existence of extremal event horizon. The above relations reduce to the limit of extremal Reissner-Nordstr\"om black hole for $b^2\rightarrow \infty$  (i.e. the limit of Maxwell's electromagnetic field theory). Using Eqs.~(\ref{eq:ge}), (\ref{eq:A_t2}) in Eq.~(\ref{eq:rdot}), we get

\begin{equation}
\dot{r}^2 = \frac{1}{m^2}\left[E-\frac{qQ}{r}\,{}_2F_{1}\left(\frac{1}{2},\frac{1}{4};\frac{5}{4};-\frac{Q^2}{b^2r^4}\right)\right]^2-g_e(r).
\label{eq:rdot2}
\end{equation}
To satisfy the condition $\dot{r}^2>0$ for the existence of the horizon radius, $r=r_e$ (Eq.~(\ref{eq:extremal_horizon2})),
\begin{equation}
E> \frac{q}{\sqrt{1-\frac{1}{4b^2Q^2}}}\cdot\,{}_2F_1\left(\frac{1}{2},\frac{1}{4};\frac{5}{4};-\frac{1}{b^2Q^2\left(1-\frac{1}{4b^2Q^2}\right)^2}\right).
\label{eq:infall2}
\end{equation}
The condition for exceeding the extremality of the final black hole becomes (by using Eq.~(\ref{eq:exceeding_extreem}))
\begin{equation}
E< \bar{Q}(Q+q,b^2) -\bar{Q}(Q,b^2),
\label{eq:exceed_extreem2}
\end{equation}
where  $M=\bar{Q}(Q,b^2)$ for the initial extremal black hole configuration and $\bar{Q}(Q,b^2)$ is given by Eq.~(\ref{eq:extremal_mass2}).

Both Eqs. (\ref{eq:infall2}) and (\ref{eq:exceed_extreem2}) will be simultaneously satisfied when $\Delta>0$ where
\begin{equation}
\Delta=\bar{Q}(Q+q,b^2) -\bar{Q}(Q,b^2) -\frac{q}{\sqrt{1-\frac{1}{4b^2Q^2}}}\cdot\,{}_2F_1\left(\frac{1}{2},\frac{1}{4};\frac{5}{4};-\frac{1}{b^2Q^2\left(1-\frac{1}{4b^2Q^2}\right)^2}\right).
\end{equation}
Assuming a small deviation from Maxwell's theory or large black hole charge such that $b^2Q^2\gg 1 $, and small charge of the test particle such that $Q\gg q$, we get
\begin{equation}
\Delta\simeq -\frac{q^2}{40b^2Q^3}.
\end{equation}
We note that overcharging of the extremal black hole is not possible for any large $b$ and small $q$. To show this for any arbitrary $b^2> \frac{1}{4Q^2} $   and $q<Q$, we define the dimensionless variable $\xi(\nu,\eta)$, (where $\nu=\frac{1}{bQ}$ and $\eta=\frac{q}{Q}$), as
 
\begin{eqnarray}
\xi = \frac{\Delta}{Q}&=& \frac{(1+\eta)}{3}\left[\sqrt{1-\frac{\nu^2}{4(1+\eta)^2}}+\frac{2}{\sqrt{1-\frac{\nu^2}{4(1+\eta)^2}}}\,{}_2F_1\left(\frac{1}{2},\frac{1}{4};\frac{5}{4};-\frac{\nu^2(1+\eta)^2}{\left((1+\eta)^2-\frac{\nu^2}{4}\right)^2}\right)\right]\nonumber\\
&& -\frac{1}{3}\left[\sqrt{1-\frac{\nu^2}{4}}+\frac{2}{\sqrt{1-\frac{\nu^2}{4}}}\,{}_2F_1\left(\frac{1}{2},\frac{1}{4};\frac{5}{4};-\frac{\nu^2}{\left(1-\frac{\nu^2}{4}\right)^2}\right)\right]\nonumber\\
&&-\frac{\eta}{\sqrt{1-\frac{\nu^2}{4}}}\,{}_2F_1\left(\frac{1}{2},\frac{1}{4};\frac{5}{4};-\frac{\nu^2}{\left(1-\frac{\nu^2}{4}\right)^2}\right).
\label{eq:xi2}
\end{eqnarray}

\begin{figure}[!htbp]
  \centering
    \includegraphics[width=3.0in,angle=360]{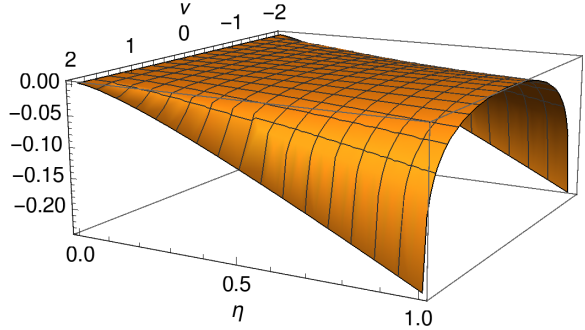}
   \caption{Surface plot of $\xi(\nu, \eta)$ for $ 4b^2Q^2\geq 1$ and $q\leq Q$. $\nu$ and $\eta$ are dimensionless variables defined as $\nu=\frac{1}{bQ}$ and $\eta=\frac{q}{Q}$. $\xi<0$ for all $\nu$ and $\eta$. Thus overcharging of an extremal black hole is not possible for any $b^2$ and $q$.}
\label{fig:geonic}
\end{figure}

We plot (3D surface plot) $\xi$ as a function of $\nu$ and $\eta$ in Fig.~\ref{fig:geonic}. The values of $\nu$ and $\eta$ are in the ranges of $-2<\nu<2$ and $0<\eta<1$. We note that $\xi$ is always negative ($\xi<0$). Thus there is no window for choosing the energy $E$ of the charged test particle such that the conditions given by Eqs.~(\ref{eq:infall2}) and (\ref{eq:exceed_extreem2}) will be simultaneously satisfied. Hence, the overcharging of an extremal geonic black hole is never possible.  This is basically an illustration of the general result obtained in  \cite{wald1974} for matter described by Born-Infeld Electrodynamics. \\

In the next subsection, we consider the case when the BI electrodynamics is coupled to EiBI gravity. Thus both the matter and the gravitational sectors are modified in accordance with the Born-Infeld prescription.

\subsection{EiBI gravity}

In EiBI gravity, the resulting black hole solutions are characterized by BI parameters, both $\kappa$ (for EiBI gravity) and $b^2$ (for BI electrodynamics) in addition to the black hole charge $Q$ and mass $M$. For detailed description of the spacetime solutions and their properties see Ref.~\cite{jana2}. Using these solutions, here, we show that overcharging of extremal black holes is possible only for a certain choice of $\kappa$ and $b^2$, particularly for $4\kappa b^2>1$. Interestingly for the case of $4\kappa b^2=1$, the conditions for choice of $E$ become exactly same as in the Reissner-Nordstr\"om black holes. Therefore it is the critical choice for $\kappa$ and $b^2$. We will analyze different situation depending on the values of $ 4 \kappa b^2$.

\subsubsection{$4\kappa b^2=1$:}
For $4\kappa b^2=1$, the metric functions take simple forms which are given by \cite{jana2} $g_{tt}=-h(r)$ and $g_{rr}=\tilde{\psi}^2(r)/h(r)$ where
\begin{eqnarray}
\tilde{\psi}&=& \left[\frac{2r^2}{r^2+\sqrt{r^4+4\kappa Q^2}}\right]^{1/2},
\label{eq:psi_tilde}\\
h(r)&=&\left(1+\frac{4\kappa Q^2}{\left(r^2+\sqrt{r^4+4\kappa Q^2} \right)^2}\right)\left[1-\frac{2\sqrt{2}M}{\sqrt{r^2+\sqrt{r^4+4\kappa Q^2}}}+\frac{2Q^2}{r^2+\sqrt{r^4+4\kappa Q^2}} \right].
\label{eq:h}
\end{eqnarray}
The spacetime looks simpler when we use a radial coordinate transformation given by 
\begin{equation}
\bar{r}=\frac{r}{\sqrt{2}}\left[1+\sqrt{1+\frac{4\kappa Q^2}{r^4}}\right].
\label{eq:rbar}
\end{equation}
Then the spacetime becomes \cite{jana2}
\begin{equation}
ds^2=U(\bar{r})\left[-\left( 1-\frac{2M}{\bar{r}}+\frac{Q^2}{\bar{r}^2} \right)dt^2+\frac{d\bar{r}^2}{\left( 1-\frac{2M}{\bar{r}}+\frac{Q^2}{\bar{r}^2} \right)}\right]+V(\bar{r})\bar{r}^2\left(d\theta^2+\sin^2\theta d\phi^2\right),
\label{eq:phyline_alpha1}
\end{equation}
where
\begin{equation}
V=1-\frac{\kappa Q^2}{\bar{r}^4}, \quad~ U=1+\frac{\kappa Q^2}{\bar{r}^4}.
\label{eq:UV}
\end{equation}
Note that $r^2=V(\bar{r})\bar{r}^2$. The spacetime (Eq.~(\ref{eq:phyline_alpha1})) resembles the Reissner-Nordstr\"om spacetime apart from the conformal factors $U$ and $V$. As $\kappa\rightarrow 0$ (and consequently $b^2\rightarrow \infty$ as $4\kappa b^2=1$) the spacetime reduces to Reissner-Nordstr\"om spacetime. There is a point singularity at $r=0$, at the location of the charge $Q$ and mass $M$.

From the equation of motion for the scalar potential $A_t$, or alternatively from the conservation of the stress-energy tensor given in Eq.~(\ref{eq:stress-tensor-BI}) (i.e. $\nabla_{\mu}T^{\mu\nu}=0$), we get
\begin{eqnarray}
\frac{dA_t}{d\bar{r}}&=& \frac{QU(\bar{r})}{\sqrt{V^2(\bar{r})\bar{r}^2+4\kappa Q^2}}
= \frac{Q}{\bar{r}^2}.
\label{eq:dA_t}
\end{eqnarray}
Thus the scalar potential $A_t$ becomes
\begin{equation}
A_t=-\frac{Q}{\bar{r}}=-\frac{\sqrt{2}Q}{r\left[1+\sqrt{1+\frac{4\kappa Q^2}{r^4}}\right]^{1/2}}.
\label{eq:A_t3}
\end{equation}
For the extremal black holes, the horizon radius $r_e$ and the relation between $Q$ and $M$ are obtained (using $h(r_e)=h'(r_e)=0$ and Eq.~(\ref{eq:rbar})) as 
\begin{eqnarray}
r_e&=&Q\sqrt{1-\frac{\kappa}{Q^2}}
\label{eq:extremal_horizon3}\\
\mbox{and}~\quad~ M&=& Q.
\label{eq:extremal_mass3}
\end{eqnarray}
Note that, for non-extremal black holes, the event horizon ($r_{+}$) and the Cauchy horizon ($r_{-}$) are given by $r_{\pm}=\left(M\pm \sqrt{M^2-Q^2}\right)\sqrt{1-\frac{\kappa Q^2}{\left(M\pm \sqrt{M^2-Q^2}\right)^4}}$.

Using Eqs.~(\ref{eq:psi_tilde}), (\ref{eq:h}), and (\ref{eq:A_t3}) in Eq.~(\ref{eq:rdot}), we get
\begin{equation}
\dot{r}^2 = \tilde{\psi}^2\left[\frac{1}{m^2}\left(E-\frac{\sqrt{2}Qq}{r}\left[1+\sqrt{1+\frac{4\kappa Q^2}{r^4}}\right]^{-1/2}\right)^2-h(r)\right].
\label{eq:rdot3}
\end{equation}

To satisfy the condition $\dot{r}^2>0$ at the horizon radius, $r=r_e$ (Eq.~(\ref{eq:extremal_horizon3})),
\begin{equation}
E> \frac{\sqrt{2}Qq}{r_e}\left[1+\sqrt{1+\frac{4\kappa Q^2}{r^4_e}}\right]^{-1/2}= q.
\label{eq:infall3}
\end{equation}

The condition for exceeding the extremality of the final black hole becomes (by using Eq.~(\ref{eq:exceeding_extreem}))
\begin{equation}
E< Q+q -M = q,
\label{eq:exceed_extreem3}
\end{equation}
where we used $Q=M$ for the initial extremal  configuration. $E>q$ and $E<q$ can not be satisfied simultaneously. We encountered exactly similar situation for extremal Reissner-Nordstr\"om black holes. Thus the overcharging of extremal black holes are not possible for any $\kappa$ provided $b^2=\frac{1}{4\kappa}$. 

\subsubsection{$4\kappa b^2>1$ :}
For $4\kappa b^2>1$, the resulting spacetime is given by \cite{jana2} $g_{tt}=-U_{\alpha}(\bar{r})h_{\alpha}(\bar{r})$ and  $g_{rr}=\frac{V_{\alpha}(\bar{r})}{U_{\alpha}(\bar{r})h_{\alpha}(\bar{r})}$ where

\begin{eqnarray}
h_{\alpha}(\bar{r})&=&1+\frac{\alpha \bar{r}^2}{6\kappa (\alpha -1)}\left[\sqrt{1-\frac{4\kappa Q^2(\alpha -1)}{\alpha \bar{r}^4}}-1\right]
 \nonumber\\
&& +\frac{\alpha^{1/4}(4Q^2)^{3/4}}{3\kappa^{1/4}(\alpha -1)^{1/4}\bar{r}}F\left(\arcsin\left(\frac{\left(4\kappa Q^2(\alpha -1)\right)^{1/4}}{\alpha^{1/4}\bar{r}}\right) \middle \vert -1\right)-\frac{2M}{\bar{r}},\label{eq:h_a}\\
U_{\alpha}(\bar{r})&=&\frac{2-\alpha}{2(1-\alpha)}-\frac{\alpha}{2(1-\alpha)}\frac{1}{\sqrt{1+\frac{4\kappa Q^2(1-\alpha)}{\alpha \bar{r}^4}}},\label{eq:U_alpha}\\
V_{\alpha}(\bar{r})&=& \frac{2-\alpha}{2(1-\alpha)}-\frac{\alpha}{2(1-\alpha)}\sqrt{1+\frac{4\kappa Q^2(1-\alpha)}{\alpha \bar{r}^4}},\label{eq:V_alpha}\\
\mbox{and}\quad~ \bar{r}&=&r\left[1-\frac{\alpha}{2}+\frac{\alpha}{2}\sqrt{1+\frac{4\kappa Q^2}{\alpha r^4}} \right]^{1/2}, \label{eq:rbar_alpha}
\end{eqnarray}
where, $F(\phi\vert m)=\int^{\phi}_0[1-m\sin^2\theta]^{-1/2}d\theta$ is the incomplete elliptic integral of the first kind and $\alpha=4\kappa b^2$. There are point singularities ($r_0=0$) for $1<\alpha\leq 2$ and surface singularities at $r_0=\left[\frac{(\alpha -2)}{2(\alpha -1)}\sqrt{\frac{4\kappa Q^2(\alpha -1)}{\alpha}}\right]^{1/2}$ for $\alpha>2$.

From the equation of motion for the scalar potential $A_t$, we get
\begin{equation}
A_t(r)= - \int_r^{\infty} \frac{Q}{x^2}\left[\left(1+\frac{Q^2}{b^2 x^4}\right)\left(1-\frac{\alpha}{2}+ \frac{\alpha}{2} \sqrt{1+\frac{Q^2}{b^2 x^4}}\right)\right]^{-1/2} dx.
\label{eq:A_t4_integral}
\end{equation} 

The last integration can be performed analytically after using the transformation $z=\sqrt{1+\frac{Q^2}{b^2 x^4}}$. We obtain
\begin{eqnarray}
A_t(r) &=& \frac{b^2}{2Q}\left(\frac{Q}{b}\right)^{3/2}\int_z^1 \frac{dz}{(z+1)^{3/4}(z-1)^{3/4}\sqrt{1+\frac{\alpha}{2}(z-1)}} \nonumber\\
&=& -\frac{\sqrt{2}b^2}{Q}\left(\frac{Q}{b}\right)^{3/2} \frac{(z-1)^{1/4}}{(2+\alpha (z-1))^{1/4}}\; {}_2F_1\left(\frac{1}{4},\frac{3}{4};\frac{5}{4};\frac{(\alpha-1)(z-1)}{2+\alpha (z-1)}\right)\nonumber\\
&=& -\frac{\sqrt{2}Q}{r}\left[\left(1+\sqrt{1+\frac{Q^2}{b^2r^4}}\right)^2+(\alpha-1)\frac{Q^2}{b^2r^4}\right]^{-1/4} \nonumber \\
&& \times {}_2F_1\left(\frac{1}{4},\frac{3}{4};\frac{5}{4};\frac{(\alpha-1)\frac{Q^2}{b^2r^4}}{\left(1+\sqrt{1+\frac{Q^2}{b^2r^4}}\right)^2+(\alpha-1)\frac{Q^2}{b^2r^4}}\right)
\label{eq:A_t4}
\end{eqnarray}

For extremal black holes, we obtain the horizon radius $r_e$, the corresponding value of $\bar{r}_e$, and the relation between $Q$ and $M$

\begin{eqnarray}
r_e &=& Q\sqrt{1-\frac{1}{4b^2Q^2}},~\quad~ \bar{r}_e= Q\sqrt{1+\frac{\alpha -1}{4b^2Q^2}},
\label{eq:extremal_horizon4}\\
M &=& \frac{\bar{r}_e}{2}\left[1+\frac{2b^2 \bar{r}_e^2}{3(\alpha -1)}\left( \sqrt{1-\frac{Q^2(\alpha -1)}{b^2\bar{r}_e^4}}-1\right) \right. \nonumber \\
&& \left. +\frac{4\sqrt{b}Q^{3/2}}{3(\alpha -1)^{1/4}\bar{r}_e}F\left(\arcsin\left(\left(\frac{Q^2(\alpha -1)}{b^2 \bar{r}_e^4}\right)^{1/4}\right) \middle \vert -1\right) \right] \nonumber\\
& =& \bar{Q}(Q,b^2,\alpha).
\label{eq:extremal_mass4}
\end{eqnarray}

To satisfy the condition $\dot{r}^2>0$ at the horizon radius, $r=r_e$ (Eq.~(\ref{eq:extremal_horizon4})),
\begin{equation}
E> q \vert A_t(r_e)\vert,
\label{eq:infall4}
\end{equation}
where $A_t(r_e)$ is to be evaluated using Eq.~(\ref{eq:A_t4}).

For exceeding the extremality of the final black hole becomes (by using Eq.~(\ref{eq:exceeding_extreem}))
\begin{equation}
E< \bar{Q}(Q+q,b^2,\alpha) -\bar{Q}(Q,b^2,\alpha),
\label{eq:exceed_extreem4}
\end{equation}
where  $M=\bar{Q}(Q,b^2,\alpha)$ for initial extremal configuration and $\bar{Q}(Q,b^2,\alpha)$ is given by Eq.~(\ref{eq:extremal_mass4}).

Both Eqs. (\ref{eq:infall4}) and (\ref{eq:exceed_extreem4}) will be simultaneously satisfied when $\Delta>0$ where
\begin{equation}
\Delta=\bar{Q}(Q+q,b^2,\alpha) -\bar{Q}(Q,b^2,\alpha) -q \vert A_t(r_e)\vert.
\label{eq:delta_4}
\end{equation}

For small deviation from Maxwell's theory or large black hole charge such that $b^2Q^2\gg 1 $ we get
\begin{eqnarray}
A_t(r_e) \simeq 1-  \frac{\alpha -1}{40b^2Q^2},\\
\bar{Q}(Q,b^2,\alpha)\simeq Q + \frac{\alpha -1}{40b^2Q},
\end{eqnarray}
where we carefully expanded all the terms in Eqs.~(\ref{eq:A_t4_integral}), (\ref{eq:extremal_horizon4}), and (\ref{eq:extremal_mass4}) up to the order $\mathcal{O}\left(\frac{1}{b^3Q^3}\right)$.

Using the above approximate results in Eq.~(\ref{eq:delta_4})  and assuming small test charge $q<< Q$, we obtain

\begin{equation}
\Delta  \simeq \frac{(\alpha -1)q^2}{40b^2Q^3} >0 .
\end{equation}
Since $\alpha=4\kappa b^2>1$, $\Delta>0$ and there is a window for choosing $E$ suitably for any small deviation from Maxwell's electromagnetic field theory.  

\begin{figure}[!htbp]
\centering
\subfigure[$\nu=1.0$]{\includegraphics[width=3.0in,angle=360]{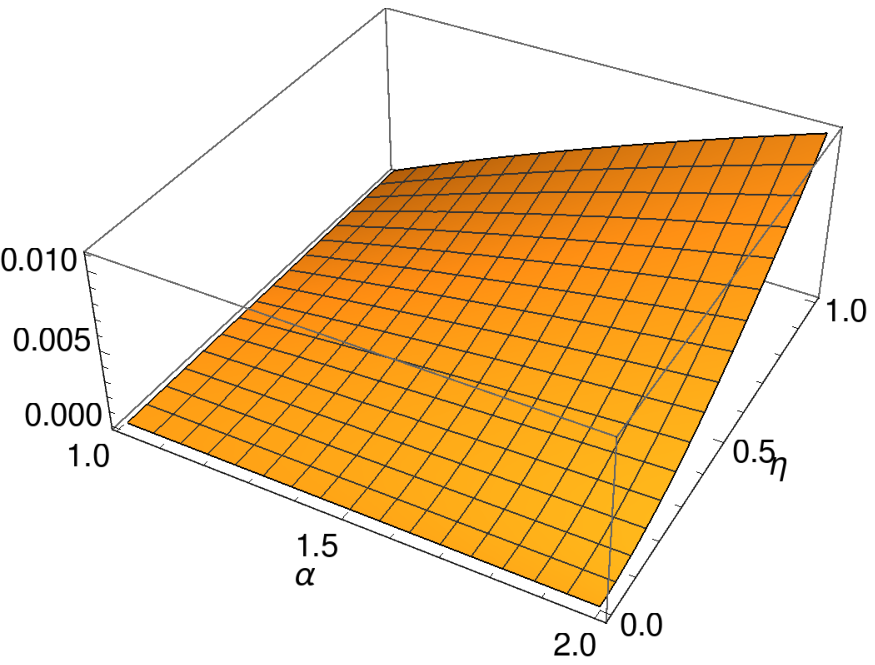}\label{subfig:eibi_maxwell1}}
\subfigure[$\nu=0.1$]{\includegraphics[width=3.0in,angle=360]{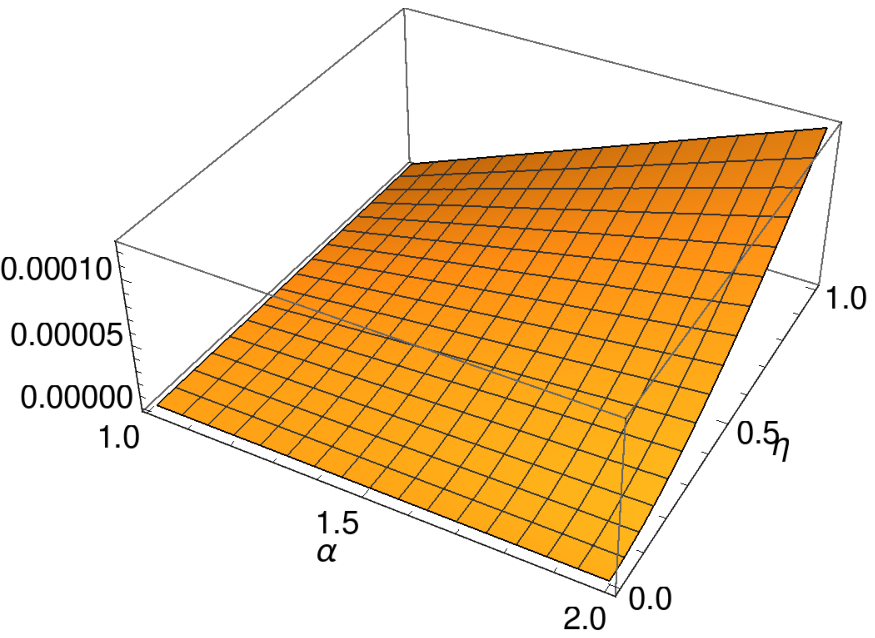}\label{subfig:eibi_maxwell2}}
\caption{Surface plot of $\xi(\eta, \alpha; \nu)$ for $q\leq Q$ and $1<\alpha\leq 2$ for the given values of $b$ ($4b^2Q^2> 1$). 
$\alpha$, $\eta$, and $\nu$ are dimensionless variables defined as $\alpha=4\kappa b^2$, $\eta=\frac{q}{Q}$, and $\nu=\frac{1}{b Q}$. In (a) $\nu=1.0$ and in (b) $\nu=0.1$. $\nu$ $\xi>0$ for all $\alpha>1$. Thus overcharging an extremal black hole is possible for $4\kappa b^2>1$.}
\label{fig:eibi_bied}
\end{figure}

To show the validity of the above result for any $b^2Q^2> 1/4$, we define a dimensionless function $\xi(\eta,\alpha;\nu)=\Delta/Q$ where $\eta=q/Q$ and $\nu=1/{bQ}$. In the Fig.~\ref{fig:eibi_bied}, we plot (3D surface) $\xi$ for two choices of $\nu$ and we note that $\xi>0$ for $4\kappa b^2 >1$ given any value of $q<Q$. Therefore both analytical and numerical analysis confirm that overcharging of an extremal black hole is possible when only $4\kappa b^2 > 1$.

\section{Conclusions}
We summarize our results point wise below:

\begin{itemize}

\item We have seen that overcharging of an extremal black hole is possible in EiBI gravity sourced by a Maxwell's electric field with black hole charge $Q$ and mass $M$. The theory parameter of EiBI gravity $\kappa$ appears in the inequalities for $E$ for a given $q$, where
$E$ and $q$ are energy and charge of the test particle of mass $m$, thrown radially to destroy the black hole. In fact, $\kappa$ generates an window for a viable choice
of $E$ satisfying the condition of overcharging. This is a significant departure from the case of general relativity.

\item Next, we investigate what would happen when we consider BI electric field instead of Maxwell's electric field. We use results of the spherically symmetric static solutions in EiBI gravity coupled BI electrodynamics \cite{jana2}. The solutions are characterized by two parameters-- $\kappa$ for EiBI gravity and $b^2$ for BI electrodynamics-- apart from charge $Q$ and mass $M$. $\kappa \rightarrow 0$ gives the GR  limit for gravitational sector and $b^2\rightarrow \infty $ gives  Maxwell's limit of BI electrodynamics theory. \\

(i) We took the solution for the critical case $4 \kappa b^2=1$, as this gives the simplest form of metric functions \cite{jana2}. For this, we interestingly found that
the criteria for overcharging an extremal black hole is exactly same as we see in the case of Reissner-Nordstrom solution, i.e. $E>q$ and $E<q$ . Thus
overcharging is not possible as long as $4 \kappa b^2=1$.

(ii) We also looked at geonic black hole solution. This is an old known solution in GR with Born-Infeld electric field instead of Maxwell's electric field
as the matter. This is also a limiting case of the solution for EiBI gravity coupled to BI electrodynamics with $\kappa \rightarrow 0$. Here also, we found that
overcharging is not possible. This is also an interesting result as ``GR coupled with BI electrodynamics" leads to that ``overcharging is not possible";  but ``EiBI gravity coupled with Maxwell's electrodynamics" leads to that ``overcharging is possible". 

\item  Extending our analysis further, we showed that in general overcharging of an extremal black hole is possible only for the case $4\kappa b^2 >1$. All of the above results are included in this inequality. 
\end{itemize}

There are several observational and theoretical justification to look for physics beyond general relativity. EiBI gravity is a viable candidate for such a modified theory of gravity. But, a modified theory of gravity is also expected to be as well behaved as Einstein's theory of general relativity. Analyzing the applicability of the Cosmic Censorship conjecture in terms of overcharging an extremal black hole solution is therefore a good consistency check for an alternative theory of gravity. In this work, we show that for the parameter range  $4\kappa b^2 >1$, such a overcharging is possible with test particle. This is completely different from the case of general relativity. As a result, it seems that the  validity of the Cosmic Censorship limits the choice of BI parameters to $4\kappa b^2 \leq 1$. Similar bounds of the parameters of a modified gravity theory like Einstein Gauss Bonnet gravity has been found using the validity of the classical second law for black holes \cite{Bhattacharjee:2015qaa}. Therefore, it may be interesting to understand further consequences of the bound $4\kappa b^2 \leq 1$ for black hole mechanics in EiBI gravity.

\section*{Acknowledgement}
Research of SS is supported by the Department of Science and Technology, Government of India under the SERB Fast
Track Scheme for Young Scientists (YSS/2015/001346).

\bibliographystyle{apsrev4-1}
\bibliography{reference}
\end{document}